\documentclass[prb,twocolumn,showpacs,preprintnumbers,amsmath,amssymb,superscriptaddress]{revtex4-1}

\usepackage{graphicx}
\usepackage{dcolumn}
\usepackage{bm}
\usepackage{psfrag} 		
\usepackage{natbib}
\usepackage{textcomp} 		
\usepackage{amsfonts}
\usepackage{amsmath}
\usepackage{color}

\newcommand{\unit}[2]{\ensuremath{#1 \, \text{#2}}}

\begin{document}



\title{Interface-induced \textit{d}-wave pairing}

\author{C.~Stephanos}
\author{T.~Kopp}
\affiliation{Center for Electronic Correlations and Magnetism, Institute of Physics, University of Augsburg, 86135 Augsburg, Germany}
\author{J.~Mannhart}
\affiliation{Center for Electronic Correlations and Magnetism, Institute of Physics, University of Augsburg, 86135 Augsburg, Germany}
\affiliation{Max Planck Institute for Solid State Research, 70569 Stuttgart, Germany}
\author{P.~J.~Hirschfeld}
\affiliation{Department of Physics, University of Florida, Gainesville, Florida 32611, USA}

\date{\today}


\begin{abstract}
We discuss a  scenario for  interface-induced  superconductivity involving pairing by dipolar excitations proximate to a two-dimensional electron system controlled by a transverse electric field.  If the interface consists of transition metal oxide materials, the repulsive on-site Coulomb interaction is typically strong and a superconducting state is formed
  via exchange of non-local dipolar excitations  in the $d$-wave channel.   Perspectives to enhance the superconducting transition temperature
 are discussed.
\end{abstract}

\maketitle

{\it Introduction.} Enormous progress has been made in recent years in the fabrication and control of interfaces of strongly correlated oxide systems.  As is well known, these materials exhibit  unusual electronic properties already in their bulk form, and the  interface of two unlike oxides appears to allow for the formation of additional novel states.  The sensitivity of such states to external parameters raises the intriguing possibility of precise quantum control of devices based on interfaces of this type. In particular, field effect devices have been  used in LaAlO$_3$/SrTiO$_3$-heterostructures to switch on and off a two-dimensional (2D) electron liquid localized at the interface~\cite{Thiel2006}. 
Recently, it was discovered that a superconducting state can be created and electrostatically modulated in such systems~\cite{Reyren2007}, albeit at quite low temperatures.  This discovery has once again raised the question of whether or not superconductivity can be influenced by interface phenomena, and indeed whether interfaces with semiconducting or insulating materials might constitute an entirely new mechanism for realizing a high-temperature superconductor. Recent discoveries of superconductivity with transition temperatures near \unit{60}{K} in Fe-based materials~\cite{FeSC} have reinvigorated discussions of novel ways to create higher temperature superconductivity, including many intriguing ideas which were effectively abandoned after they went unrealized in the early years after the advent of BCS theory.  In this paper, we discuss a mechanism for interface-mediated superconductivity specific to oxide interfaces and investigate whether high critical temperatures might be possible.

Our discussion is very much in the spirit of an early model due to Little, who  proposed that high temperature superconductivity with a temperature scale determined by electronic energy scales of the order of the Fermi energy might be realized with  metalorganic chain compounds with quasi-1D metallic spines coupled to polarizable (organic) side chains~\cite{Little1964}. The electronic excitations in the side chains were assumed to induce   Cooper pairing in the metallic spine. A crucial element of Little's scheme was the spatial separation of the metallic electrons from the excitations mediating pairing. This separation prevented the screening of the excitations which otherwise would have lowered the pairing scale. In fact, Little and Gutfreund argued that exchange effects and vertex corrections associated with interactions between metallic carriers and the excitations in the side chains do not spoil  Cooper-pair formation if the separation is sufficiently large~\cite{Gutfreund1979}. These effects are expected to decay exponentially with distance,  whereas the  Coulomb potential between polarizable side chains and charge carriers in the spine relevant for pairing depends algebraically on the respective distance. Similar concepts were elaborated for two-dimensional interfaces between metals and dielectrics by Ginzburg~\cite{Ginzburg1964};  here one expects that, although true long-range order is not allowed by the Mermin-Wagner theorem, algebraic order at finite temperatures is still feasible. Allender, Bray, and Bardeen proposed an excitonic mechanism in superconductor-semiconductor sandwiches~\cite{Allender1973}. In this scheme, the exponential tails of the electronic wave functions from the metallic side experience an effective pairing  potential mediated through the particle-hole excitations across the semiconductor gap. They predicted a superconducting state at the metallic surface which, under favorable conditions, might persist to a critical temperature $T_\text{c}$ of the order of several tens of Kelvin.

\begin{figure}[htbp]
\centering
\includegraphics[width=0.9\columnwidth]{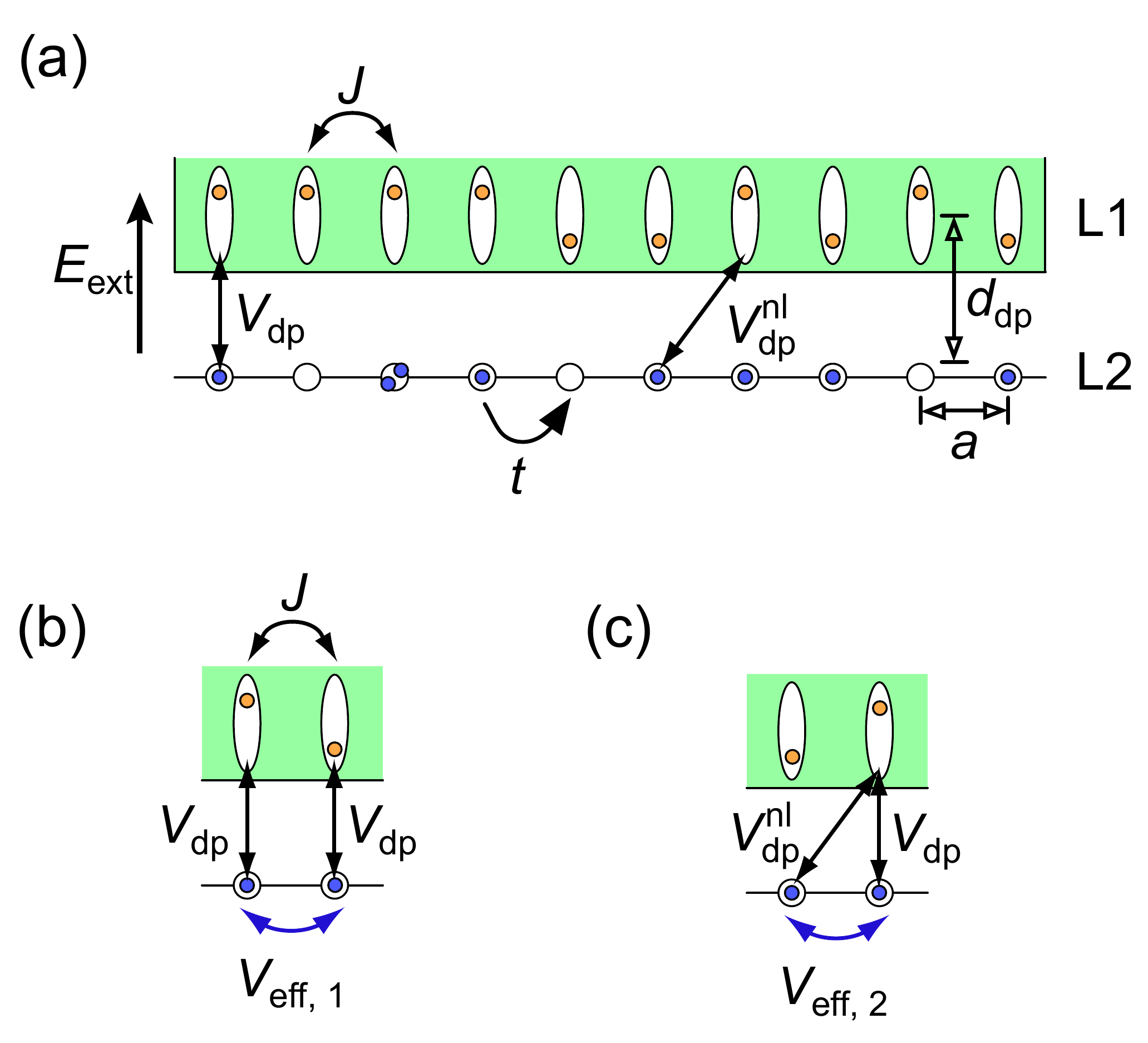}
\caption{a) Geometry of the model system and the different interaction terms as described in text; b), c) Illustration of different interaction mechanisms leading to effective non-local electron pairing.} 
\label{img:interactions}
\end{figure}

There are no superconductors to our knowledge where Cooper pairing arises via an excitonic mechanism of this general type. It is impeded by the necessity of reconciling two opposing tendencies: on the one hand, the two electronic subsystems --- the charge carriers in the metal layer and the polarizable side chains or ligands --- have to be well separated to avoid exchange effects.  On the other hand, they should be as close as possible to generate a strong pairing potential. In this respect, an idea due  to Hirsch and Scalapino appears quite promising: they propose searching for a material in which the two electronic subsystems are not well separated in space but have mutually orthogonal wave functions~\cite{Hirsch1985}. This implies that  vertex corrections to the pairing are suppressed very effectively although the subsystems reside partially on the same transition metal sites.  Hirsch and Scalapino did not find an easy route to high temperature superconductivity through models of this kind. This, however, may possibly be due to the restriction to 1D systems.

{\it Fundamental model for bilayer superconductivity.}
At interfaces of materials, electronic properties  can be dramatically different from their respective bulk values at either side of the interface.
In this work we study a mechanism for interface-mediated superconductivity induced by the interaction of two separated electron systems. The model system consists of two layers L1 and L2. Layer L2 comprises a two-dimensional electron system on a square lattice with lattice constant $a$ and nearest-neighbor hopping $t$. Adjacent to every lattice site, in a distance $r$, there is a localized two-level (2D) system in layer L1 with ground state $d$, excited state $p$ and excitation energy $\Delta_{\text{dp}}$. These excitations can be modeled as dipoles with a dipole moment $ed_{\text{dp}}$ with the constraint $d_i^\dagger d_i + p_i^\dagger p_i = 1$.
An external electric field applied perpendicular to the layer structure $E_z$ both influences the charge carrier density in the electron layer L2 and polarizes the two-level systems in L1. In the scope of this work we will refer to two-level systems in L1 adjacent to an electron lattice site in L2 as a dipole on the same lattice site. The basic model described until now is identical to the model used by Koerting \textit{et al.} \cite{Koerting2005}.

The local $s$-wave pair state found by Koerting \textit{et al.}, is suppressed by a strong on-site Coulomb repulsion between electrons. In this work, we investigate an extension to the model system by including non-local interactions:
 electrons not only interact with dipoles on the same but also with dipoles on the next lattice site, with the respective interaction energies $V_{\text{dp}}$ and $V_{\text{dp}}^{\text{nl}}$. Nearest-neighbor dipoles interact with energy $J$. Fig.\,\ref{img:interactions}a illustrates the geometry of the model system and the different interaction mechanisms leading to effective non-local pairing (Figs.\,\ref{img:interactions}b, c).

The Hamiltonian assumes the following form:
\begin{align}
H=H_{\text{kin}} + H_{\text{e-e}} + H_{\text{2l}} + H_{\text{ext}} + H_{\text{dip-dip}} + H_{\text{e-dip}}
\end{align}

with

\begin{align}
& H_{\text{kin}} = -t \sum\limits_{<i, j>, \sigma} c_{i, \sigma}^{\dagger} c_{j, \sigma} \label{H_hopping} \\
& H_{\text{e-e}} = U  \sum\limits_{i, \sigma}  n_{i, \sigma}  n_{i, -\sigma}+V^{\text{nl}} \sum\limits_{\substack{<i, j>\\ \sigma, \sigma'}} n_{i, \sigma} n_{j, \sigma'} \label{H_coul}\\
& H_{\text{2l}}  = \frac{1}{2}\Delta_{\text{dp}} \sum_i \left( p_i^\dagger  p_i - d_i^\dagger d_i \right) \label{H_tl}\\
& H_{\text{ext}} = E_{\text{ext}} \sum_i \left(p_i^\dagger d_i + d_i^\dagger p_i \right) \label{H_E} \\
& H_{\text{dip-dip}}  = -J \sum\limits_{<i, j>} \left (p_i^\dagger d_i + d_i^\dagger p_i \right)\left( p_i^\dagger d_j + d_j^\dagger p_j \right) \label{H_J}\\
& H_{\text{e-dip}} = V_{\text{dp}} \sum_{i, \sigma} n_{i, \sigma} \left(p_i^\dagger d_i + d_i^\dagger p_i \right)\nonumber \\  &~~~~~~~~~~~~+ V_{\text{dp}}^{\text{nl}} \sum_{<i, j>, \sigma} n_{i, \sigma} \left(p_j^\dagger d_j + d_j^\dagger p_j \right) \label{H_int}
\end{align}
where $n_{\sigma} = c_{i, \sigma}^\dagger c_{i, \sigma}$ and $E_\text{ext} = e d_\text{dp} E_z $.
$H_{\text{kin}}$ is the Hamiltonian for a band of non-interacting electrons in a 2D square lattice, $H_{\text{e-e}}$ represents the Coulomb repulsion between electrons on the same (first part) and on nearest-neighbor (second part) lattice sites, $H_{\text{2l}}$ the localized two-level systems (dipoles), $H_{\text{ext}}$ the coupling of the electric field to the dipoles, and $H_{\text{dip-dip}}$ the nearest-neighbor dipole interaction. $H_{\text{e-dip}}$ is the interaction between electrons in the metallic layer L2 and dipoles in layer L1. The first part describes a local interaction, whereas the second part represents a non-local interaction between electrons and dipoles on nearest-neighbor lattice sites (see Fig.\,\ref{img:interactions} a--c for the different interaction terms).

Following \cite{Koerting2005}, we diagonalize $H_{\text{2l}}  + H_{\text{ext}}$ in a first step and replace the quasi-particle operators by a pseudospin representation, where the longitudinal component $S_i^\text{z}$ measures the occupation of the two-level system and the transverse components $S_i^\pm$ induce transitions between two energy states. We then apply a Holstein-Primakoff transformation~\cite{Holstein1940} on bosonic variables ($b_j$, $b_j^\dagger$) and a Lang-Firsov transformation~\cite{Lang1963} (LFT). To identify the terms controlling the non-local pairing interaction between electrons on next-neighbor lattice sites, we carry out a modified version of the LFT. The non-local expansion of the LFT allows us to eliminate terms of the form $\sum\limits_{<i,j>, \sigma} \!\!\! n_{i, \sigma}(b_j + b_j^\dagger)$, generating a direct interaction term 
$V_\text{eff}\!\!\! \sum\limits_{<i, j> \sigma,\sigma'} \!\!\! n_{i, \sigma} n_{j, \sigma'}$ 
between two electrons on nearest-neighbor lattice sites with the effective interaction energy $V_\text{eff}$.  After expanding, assuming $E_\text{ext} \ll V_\text{dp} < \Delta_\text{dp}$ and $J < \Delta_\text{dp}$ and performing a Bogoliubov transformation~\cite{Koerting2005}, we find 
$
V_\text{eff} = V_\text{eff, 1} + V_\text{eff, 2} - V_\text{E} - V_\text{e-e},
$
where
$
V_\text{eff, 1} = J \frac{\Delta_\text{dp}^2}{\lambda^2} g^2$,  $V_\text{eff, 2} = 2 V_x^\text{\rm nl} g$, and $ V_\text{E} = V_z^\text{nl} g^2$. Here we have defined $V_x=V_{\rm dp}\Delta_{\rm dp}/(2 \lambda)$, $V_x^{\rm nl}=V_{\rm dp}^{\rm nl}/(2 \lambda)$ and $\lambda = \sqrt{E_\text{ext}^2 + \frac{1}{4}\Delta^2_\text{dp}}$.
The first part $V_\text{eff, 1}$ describes the effective electron-electron pairing induced by the nearest-neighbor coupling of the dipoles (Fig\,~\ref{img:interactions}b). The second part $V_\text{eff, 2}$ originates from electrons interacting with dipoles on the next lattice sites (Fig.\,\ref{img:interactions}c). $V_\text{E}$ is repulsive, caused by  the polarization of the dipoles due to the electric field, analogous to the local term found by Koerting \textit{et al.}~\cite{Koerting2005}. The nearest-neighbor Coulomb repulsion $V_\text{e-e}$  weakens the attractive interaction. The factor $g=V_x/(2\lambda V_x(1+n)+V^{\rm nl}_x n)$ introduced by the local LFT, effectively describes the interaction between an electron and a dipole on the same lattice site.

{\it Results.}
We first fix model parameters. It is beyond the scope of our  investigation to identify the various parameters for specific materials. 
For concreteness, we assume that the two layers L1 and L2 compose a perovskite heterostructure, e.g.  SrTiO$_3$ (L1) and YBa$_2$Cu$_3$O$_6$ (L2), as studied in Ref.~\onlinecite{Pavlenko2007} within a DFT approach. For the evaluation we introduce the ``standard set'' of parameter values: the lattice constant is $a =4$~\AA, the distance between layers L1 and L2 is taken to be the distance between a lattice site in L2 and the midpoint of the dipole in L1 and is fixed to $r =3.4$~\AA, and the dipole length is approximately the distance between a metal atom at the center of an octahedron in L1 and its ligand  in the direction towards the layer L2, notably $d_{\text{dp}}=1.9$~\AA. We choose a value $t=0.25$~eV for the hopping amplitude in L2, and a nearest-neighbor Coulomb repulsion of $V^{\text{nl}}=0.5 \, t$. The on-site repulsion $U$ in L2 is of no relevance for our considerations as we do not include a local $s$-wave pairing channel in the evaluation; we entirely focus on the non-local $d$-wave pairing (we have not explored possible extended-$s$ pairing at low electron densities which however has been estimated to be weaker). For the energy separation of the two-level systems in L1 we choose $\Delta_\text{dp}/4t = 2.5$ which is a typical value for charge-transfer excitations in transition metal oxides.

\begin{figure}[htbp]
\centering
\includegraphics[width=1\columnwidth]{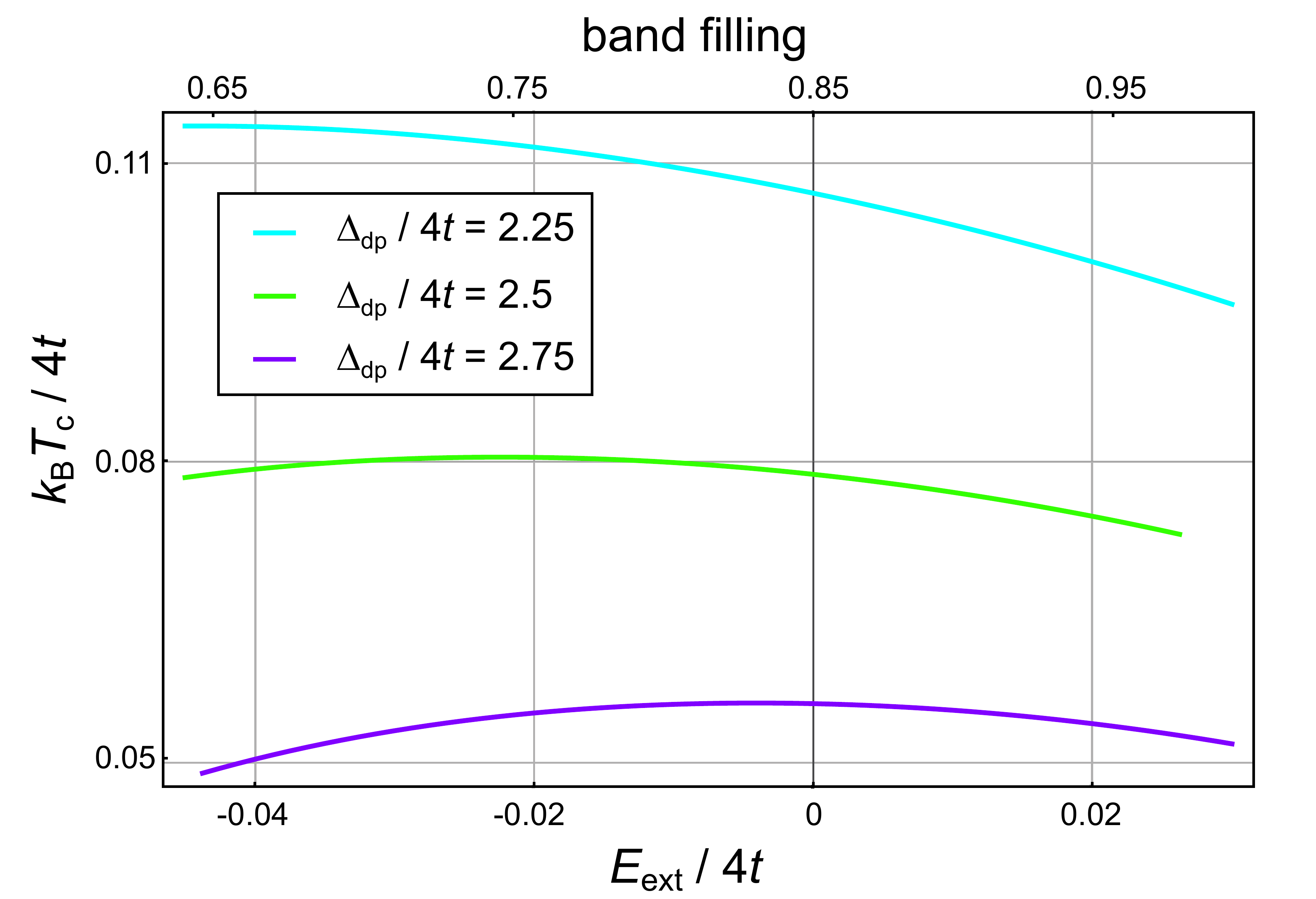}
\caption{ The critical temperature $T_\text{c}$ as criterion for the interface mediated \textit{d}-wave superconductivity is displayed as function of the external electric field energy $E_\text{ext}$ for different values of the dipole energy gap $\Delta_\text{dp}$.}
\label{img:Tc}
\end{figure}

To estimate the  interactions between charge carriers and dipoles, 
we must consider the screening of the microscopic processes  produced by local excitations (polarizations), which may be considerably smaller than implied by the bulk dielectric constant (see the discussion in Refs.~\onlinecite{Brink1997,Boer1984}). For concreteness, we choose a screening parameter $\epsilon_\text{s}=3$ for L1 which may be approximately identified from LDA-calculations for a summation of interband transitions in such transition metal oxides~\cite{Pavlenko2009}. 
We now estimate the remaining interaction parameters $V_{\text{dp}}$, $V_{\text{dp}}^{\text{nl}}$, and $J$.
The interaction energy between the dipoles next to the interface
and the electric field of a charge carrier on the nearest site is given \cite{Koerting2005} by
$
V_{\text{dp}}={e^2 d_{\text{dp}}}/({{4\pi\epsilon_0}\epsilon_\text{s} r^2})\simeq 3.1 \,t$, and
the non-local interaction between electrons and dipoles on nearest-neighbor lattice sites is analogously
$
V^{\text{nl}}_{\text{dp}}= {e^2 d_{\text{dp}}}/({{4\pi\epsilon_0}\epsilon_\text{s} (r^2 +a^2)} )\simeq 1.3\,t$.
The interaction energy between two dipoles in layer L1 is (see Appendix B of Ref.~\cite{Boer1984})
$
J= {e^2 d^2_{\text{dp}}}/({{4\pi\epsilon_0}\epsilon_\text{s} a^3})  \simeq 1.1\,t$.  
Excited electrons in L1 can also hop between adjacent centers. Virtual hopping processes of this kind generate a similar interaction term and renormalize the energy scale $J$. However we neglect this additional contribution as it does not qualitatively affect the considered pairing mechanism. 

Finally, if field doping is included through the external field $E_\text{ext}$ we have to introduce the induced interfacial electronic charge~\cite{Koerting2005}
$
n=[\epsilon_0\epsilon_\text{b} a^2 /( e^2 d_{\rm dp})]E_{\rm ext}$,
where $\epsilon_\text{b}$ is the bulk dielectric constant of the gate dielectric.

With the specification of the standard set of parameter values, we determine the filling-dependent BCS transition temperature to the $d$-wave superconducting state from the effective  interaction $V_{\rm eff}$ above. For Fig.\,\ref{img:Tc} the transition temperature $T_\text{c}$ has been calculated for an $E_\text{ext}=0$  band filling of $n=0.85$ for three values of the excitation energy $\Delta_{\text{dp}}$. For values close to that of the standard parameter set, $k_\text{B} T_\text{c}/4t$ is of the order of $10^{-1}$. Although fluctuations will suppress the high value for $T_\text{c}$, the scale is set by the non-local exchange of excitations in layer L1 (cf.~Figs.\,\ref{img:interactions}b and \ref{img:interactions}c). Both of the  exchange processes considered contribute with a similar strength. For virtual excitations with higher values of the excitation energy $\Delta_{\text{dp}}$, the transition temperature drops. For low values,  where $\Delta_{\text{dp}}$ is of the order of $k_\text{B} T_\text{c}$, the number of inverted two-level systems in L1 becomes sizable and $T_\text{c}$ as function of $\Delta_{\text{dp}}$ drops again; we do not consider this limit of small $\Delta_{\text{dp}}$ in this work. For $\Delta_{\text{dp}}/4t = 2.5$, the number of inverted two-level systems is of the order of $10^{-9}$.

The smooth curves in Fig.\,\ref{img:Tc} display the dependence of $T_\text{c}$ on field induced doping. The transition temperature decreases for sufficiently negative field value since $T_\text{c}$ depends  on the density of mobile charge carriers. However, $T_\text{c}$ also decreases for positive field values because the effective interaction is suppressed by high electric field strength~\cite{Koerting2005}: the term $V_\text{E}$ represents the interaction of field induced dipoles in L1 with the charge carriers in L2 and is repulsive~\cite{Koerting2005}. Here, it should be noted that $V_\text{E}$ increases linearly with the field strength $E_{z}$ for small fields ($e d_\text{dp} E_z \ll \frac{1}{2}\Delta_\text{dp}$). Moreover, $T_\text{c}$ decreases for increasing $E_z$ as the effective excitation gap of the two-level systems is increased.

\begin{figure}[htbp]
\centering
\includegraphics[width=1\columnwidth]{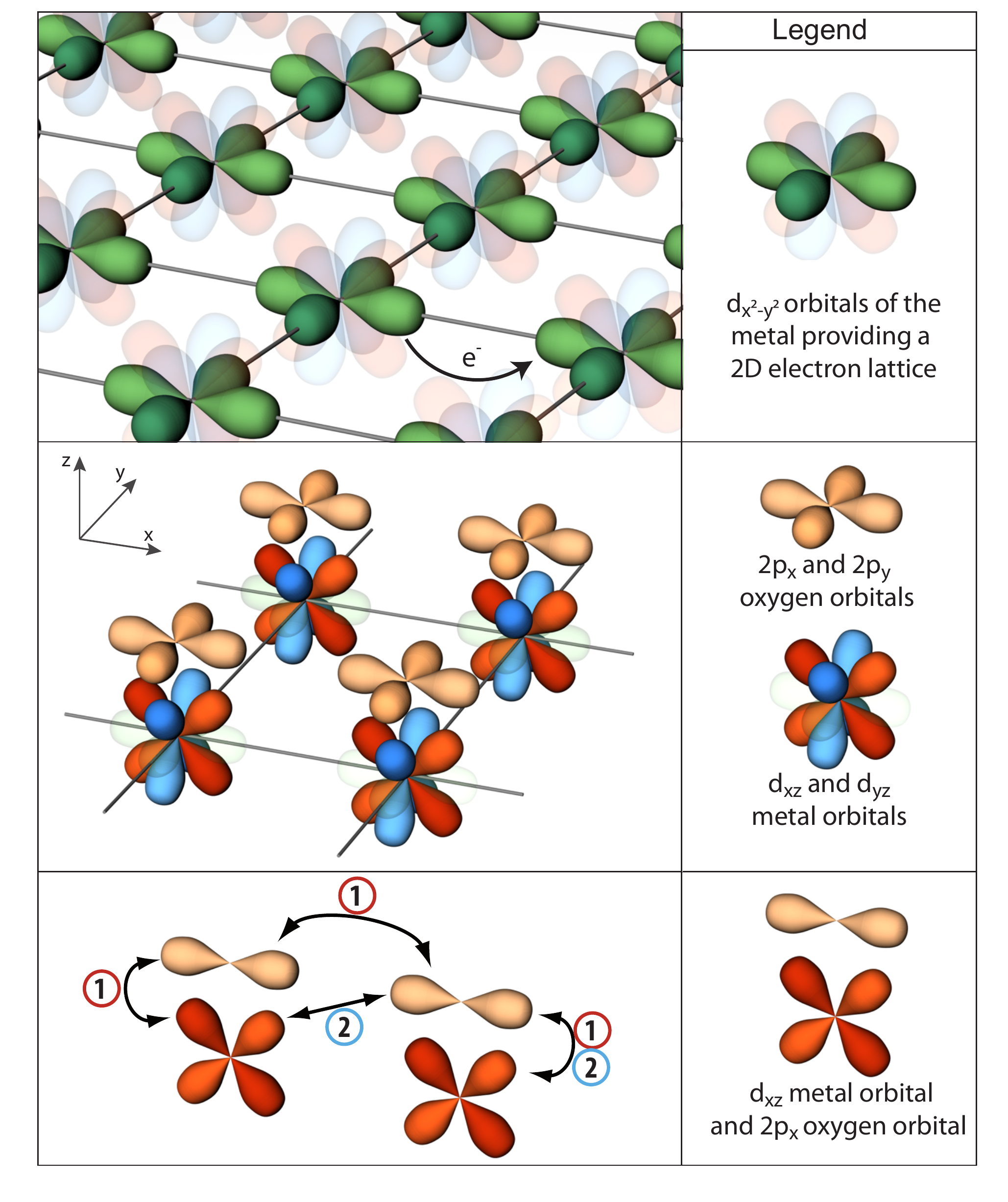}
\caption{Proposed realization of the L1/L2 model utilizing orthogonal atomic orbitals.}
\label{img:orbitals}
\end{figure}

{\it Conclusions.}
We have presented a concrete calculation of possible interface-induced $d$-wave pairing 
mediated by localized electronic excitations localized in a proximate surface layer.  The advantage of
such a geometry, as noted earlier by Little, Ginzburg, and others, is that
 the spatial separation of the excitations mediating pairing from the metallic electrons helps to prevent the screening of the excitations.  As in such early approaches, suppression of the critical temperature
by exchange effects is avoided here by this spatial separation.
The unique advantage of the current scheme over earlier excitonic 
mechanisms of this type, however, is the  avoidance of the repulsive local Coulomb interaction
$U$ through the $d$-wave pairing state realized.  The simplest calculation of 
the bare transition temperature is then on the scale of $J\cdot(V_{\text{dp}}/\Delta_\text{dp})2$, 
and can easily be of order several 
hundred K.  Of course vertex corrections will suppress this temperature substantially,
and due to the large momentum transfer in the interactions leading to d-wave
pairing, the vertex effects will not be weakened as substantially as, e.g.
in Ref. \cite{Gutfreund1979}.  Whether the first or second effect dominates
in real systems is dependent on specifics of materials we have been unable to
address here; we merely point out that new possibilities for the optimization 
of interface-induced pairing are raised by the  oxide heterostructures.

Of course, the separation of pairing excitation and metallic layers also reduces the intrinsic strength of the
pairing.  It is therefore interesting to speculate further on the scheme of Scalapino and Hirsch~\cite{Hirsch1985}, extended to the current 2D context, by imagining concatenating layers L1 and L2, and
suppressing vertex corrections by constructing the two subsystems from mutually orthogonal atomic orbitals on
the same sites, as illustrated in Fig.~\ref{img:orbitals}. In the upper panel, a conductive layer is formed by metal atoms on a square lattice, e.g. in a perovskite structure. Electrons then hop, for example, within the $d_{x^2 - y^2}$ orbitals of the metal atoms. In the central panel, the non-local electron-electron interaction is mediated, e.g., by the $d_{xz}$, $d_{yz}$ orbitals of the metal and the $p_x$ and $p_y$ orbitals of oxygen atoms, which are adjacent to every metal atom. In c) we illustrate two different interaction mechanisms, viz: an excitation can jump along either path to enable a non-local electron-electron interaction which would be similar to the one described here.

Finally, we mention that the concepts discussed here in the context of interface-mediated pairing may also be applicable to the insulating spacer layers of the cuprates in bulk.  As discussed by Eisaki {\it et al.}~\cite{Eisaki}, if the pairing interaction for cuprate superconductors arises universally in the CuO$_2$ plane, as is commonly assumed, it is difficult to understand the large range of $T_\text{c}$'s associated with the single layer cuprate materials, from 10~K to 90~K.  It is possible that excitonic pairing with a proximate spacer layer may bootstrap the basic in-plane pairing interaction in certain systems.  To enhance rather than suppress pairing, however, it must have the same symmetry as the in-plane pairing interaction, i.e., $d$-wave, just as in the mechanism discussed here.

{\it Acknowledgements.}  The authors are grateful to T.~Geballe, N.~Pavlenko,  and D.~J.~Scalapino for many inspiring discussions. CS, TK and JM were supported by the DFG through TRR~80; PJH was supported by DOE DE-FG02-05ER46236.



\bibliographystyle{aipnum4-1}

%
%
%
%
%
%
%


%

\end{document}